# Dynamic atomic reconstruction:
# how $Fe_3O_4$ thin films evade polar catastrophe for epitaxy

**Authors:**

C. F. Chang,[1] Z. Hu,[1] S. Klein,[2,†] X. H. Liu,[1] R. Sutarto,[2,†] A. Tanaka,[3] J. C. Cezar,[4,†] N. B. Brookes,[4] H.-J. Lin,[5] H. H. Hsieh,[6] C. T. Chen,[5] A. D. Rata,[1,†] and L. H. Tjeng[1]

**Affiliations:**

[1]Max Planck Institute for Chemical Physics of Solids, Nöthnitzerstr. 40, 01187 Dresden, Germany

[2]II. Physikalisches Institut, Universität zu Köln, Zülpicher Str. 77, 50937 Köln, Germany

[3]Department of Quantum Matter, ADSM, Hiroshima University, Higashi-Hiroshima 739-8530, Japan

[4]European Synchrotron Radiation Facility, 71 Avenue des Martyrs, Grenoble, France

[5]National Synchrotron Radiation Research Center, 101 Hsin-Ann Road, Hsinchu 30077, Taiwan

[6]Chung Cheng Institute of Technology, National Defense University, Taoyuan 335, Taiwan

[†] Current address;

S. Klein: Institute of Materials Physics in Space, German Aerospace Center, Linder Hohe, 51147 Köln, Germany;

R. Sutarto: Canadian Light Source, Saskatoon, Saskatchewan S7N 2V3, Canada;

J. C. Cezar: Laboratório Nacional de Luz Síncrotron, C.P. 6192, 13083-970 Campinas, SP, Brazil;

A. D. Rata: Institute for Physics, Martin-Luther-University, Halle-Wittenberg, 06099 Halle, Germany;

**Abstract**:

**Polar catastrophe at the interface of oxide materials with strongly correlated electrons has triggered a flurry of new research activities. The expectations are that the design of such advanced interfaces will become a powerful route to engineer devices with novel functionalities. Here we investigate the initial stages of growth and the electronic structure of the spintronic $Fe_3O_4$/MgO (001) interface. Using soft x-ray absorption spectroscopy we have discovered that the so-called A-sites are completely missing in the first $Fe_3O_4$ monolayer. This allows us to develop an unexpected but elegant growth principle in which during deposition the Fe atoms are constantly on the move to solve the divergent electrostatic potential problem, thereby ensuring epitaxy and stoichiometry at the same time. This growth principle provides a new perspective for the design of interfaces.**

**Subject areas**: Condensed Matter Physics, Magnetism, Strongly Correlated Materials, Materials Science, Spintronics, Chemical Physics, Nanophysics

**Main Text:**

Physical properties of surfaces and interfaces of solids could markedly differ from those in the bulk, especially in cases when the surface or interface involves non-neutral crystal planes. For insulators, these polar planes cause the electrostatic potential to diverge, and thus to destabilize the system dramatically. The surface or interface is then forced to reconstruct, for example, by forming facets or defects. Or more spectacular, it could give up a substantial amount of charge, thereby altering its electronic structure completely [1–3]. Recently, claims have been made that such happens at the $SrTiO_3$/$LaAlO_3$ interface since the interface is conducting while the two constituents separately are good insulators [4, 5]. In fact, the formation of a two-dimensional electron gas and the occurrence of superconductivity in this interface has generated frantic efforts worldwide to explore the potential of these interfaces for device applications [6–12] as well as to search for new emergent phenomena such as quantum criticality in 2-dimensional electron gas systems [13–15]. However, the growth process of these interfaces and polar interfaces in general is a mystery. How do the atoms rearrange themselves during the deposition or growth such that a well ordered and smooth interface is formed, despite the destabilizing forces due to the catastrophic electrostatic potential? Understanding the growth principles will widen the scope of interfaces that one may want to design: interfaces which appear impossible to grow at first sight may now be tried out.

Here, we investigate the polar interface between $Fe_3O_4$ and the MgO (001) substrate, one of the most used interfaces in the research field of spintronics. [16-24] This interface is completely not understood in terms of atomic structure, electronic structure and growth mode. Figs. 1 (a), and (b) illustrate how the $Fe_3O_4$ inverse spinel crystal structure consisting of $Fe^{3+}$ ions in tetrahedral coordination (A-sites), $Fe^{2+}$ and $Fe^{3+}$ ions (or on average $Fe^{2.5+}$) in octahedral coordination (B-sites) with $O^{2-}$ ions in an fcc-lattice builds up the polar catastrophe problem. We have prepared the system using molecular beam epitaxy (MBE). This deposition method allows for a layer-by-layer growth of the $Fe_3O_4$ under ultra high vacuum conditions which facilitates the use of *in-situ* characterization techniques with surface monolayer sensitivity. The MBE growth of $Fe_3O_4$ on MgO (001) is also known to produce films with excellent physical properties [22]. In order to obtain direct insight into the atomic and electronic structure of the interface, we utilize soft x-ray absorption spectroscopy (XAS) at the Fe $L_{2,3}$ edges. This spectroscopic technique is extremely sensitive to the local coordination and charge state of the Fe ions [25–28].

$Fe_3O_4$ thin films with thicknesses varying between 0.67 and 8 monolayers (ML) were grown on MgO (001). Each film has been grown on a new and freshly annealed substrate. The substrate temperature was kept at 250 °C during the growth in order to avoid the Mg inter-diffusion at the $Fe_3O_4$/MgO interface [29,30]. Details about the film growth are given in the Supplementary Materials [31]. One ML consists of one (001)-oriented layer of oxygen anions together with the appropriate number of Fe cations to maintain charge neutrality and stoichiometry, and has a thickness of 2.1 Å. In Fig. 1 (c), and (d) we present representative reflection high energy electron diffraction (RHEED) and low energy electron diffraction (LEED) patterns, respectively, of a 200 nm thick $Fe_3O_4$ film to demonstrate that the surface is still smooth for very long deposition times. The typical $(\sqrt{2} \times \sqrt{2})R45°$ surface reconstruction is also clearly visible. Fig. 1 (e) shows the regular oscillations with time in the intensity of the specularly reflected RHEED beam during growth, indicating a two-dimensional layer-by-layer growth mode. Fig. 1 (f) displays the resistivity as a function of temperature from a 10 nm and a 200 nm film, showing the presence of the characteristic Verwey transition [22].

Figure 2 depicts the room temperature Fe $L_3$ XAS spectra of $Fe_3O_4$ films with thicknesses varying from 0.67 ML to 8 ML, and of a $Fe_3O_4$ bulk single crystal. A $Fe_2O_3$ single crystal was measured simultaneously in a separate chamber to serve as energy reference for Fe $L_3$ edge. Further XAS experimental details and display of the entire Fe $L_{2,3}$ spectral range are given in the Supplementary Materials [31, Fig. S1]. We also include in Fig. 2 the spectra of bulk YBaCo$_3$FeO$_7$ [28], bulk FeO (reproduced from Ref. *32*) and bulk Fe$_2$O$_3$ as references for Fe$^{3+}$ ions in tetrahedral coordination, Fe$^{2+}$ ions in octahedral coordination, and Fe$^{3+}$ ions in octahedral coordination, respectively. The line shapes of the spectra strongly depend on the multiplet structure given by the atomic like Fe 3*d*–3*d* and 2*p*–3*d* Coulomb and exchange interactions, as well as by local crystal fields and the hybridization with the O 2*p* ligands [25–28]. Here we note the striking similarities of the spectral features of the 8 ML Fe$_3$O$_4$ thin film and the bulk magnetite, which confirms that our Fe$_3$O$_4$ films have the correct stoichiometry.

We now focus on the thickness dependence of the spectra. Clear and systematic changes can be observed, in particular in the peak position of the spectral feature labeled as (I) and in the intensity of the spectral feature labeled as (II) relative to that of peak (I), see Fig. 2. The position of peak (I) of the thinnest Fe$_3$O$_4$ films, *i.e.* of the 0.67, 0.75 and 1 ML films, is the same as that of bulk Fe$_2$O$_3$, while for the thicker films, *i.e.* 2 ML and beyond, it is more similar to that of bulk YBaCo$_3$FeO$_7$. This gives a first indication that the thinnest films contain only tiny amounts of Fe$^{3+}$ ions in tetrahedral coordination and implies that such A-site Fe ions could essentially only be present for films of 2 ML thickness and beyond. This then would also explain why for the thinnest films one can see two separate peaks (I) and (II) like in bulk Fe$_2$O$_3$ (green curve), while for thicker films the appearance of an in-between peak associated with the Fe$^{3+}$ ions in tetrahedral coordination (red curve) will fill up the valley between peak (I) and (II), making peak (II) to become a shoulder and the position of the larger peak (I) to shift to lower energies. Important is to note that the foot at the onset of the Fe $L_3$ edge, *i.e.* the feature between 706 and 707.7 eV, which is part of the spectral feature characteristic for Fe$^{2+}$ ions in octahedral B sites (see blue curve), are thickness independent. All these strongly suggest that the spectral weight of the A-site and the B-site Fe$^{3+}$ ions varies strongly with thickness.

To interpret and better understand the XAS spectra and their thickness dependence we have performed calculations using the well established configuration interaction cluster model that includes the full atomic multiplet theory and the local effects of the solid [25–28]. We have simulated each of the XAS spectra shown in Fig. 2 and obtained the spectral weight of the different Fe sites; computational details and fits are given in the Supplementary Materials [31,

Figs. S2, S3, and S4, 33]. The results are plotted as closed squares in the top panel of Fig. 3, where the error bars reflect the deviations of the fits to the experimental data. From the relative concentrations of the constituents we have calculated the average valence or equivalently, by taking the oxygen lattice to be complete, we have determined the Fe content y in our $Fe_yO$ films. These y values are plotted as black closed squares in the bottom panel of Fig. 3. We can observe that all points are very close to the $Fe_{3/4}O$ (gray) line, which confirms the correct stoichiometry of our films through the entire thickness range and very consistent with the RHEED intensity oscillations which have a constant time period, *i.e.* independent of the film thickness.

An important aspect that emerges directly from the simulations is the strong thickness dependence of the different Fe constituents, see Fig. 3. We recall that bulk $Fe_3O_4$ has 1/3 (33%) $Fe^{3+}$ ions in tetrahedral coordination (A-sites), 1/3 (33%) $Fe^{2+}$ and 1/3 (33%) $Fe^{3+}$ ions in octahedral coordination (B-sites). We found to our surprise that the amount of A-site $Fe^{3+}$ ions is practically negligible for the thinnest films, *i.e.* 2-3% instead of the 33% bulk value. At the same time, the amount of B-site $Fe^{3+}$ in the thinnest films is between 60-68%, much larger than the 33% bulk value. We also observe that with increasing film thickness the A-site $Fe^{3+}$ amount increases and the B-site $Fe^{3+}$ decreases, both to approach the 33% bulk value, see for example the 8 ML results in Fig. 3. Interestingly, the amount of B-site $Fe^{2+}$ is rather constant and independent of the film thickness, it fluctuates around the 33% bulk value.

These spectroscopic findings provide crucial data for the determination of the actual growth process and the interface structure. Especially the observation that the first monolayer of the $Fe_3O_4$ film has essentially no A-sites is a surprising piece of information. In fact, as far as the monolayer is concerned, the choice of 'nature' not to have A-sites is the simplest manner to solve the planar electrostatic potential problem. As can be seen from Figs. 1 (a) and (b), it is indeed the presence of the A-sites that causes the polar catastrophe to occur as there are no negative ions in those A-site planes to neutralize the charges. So by not having A sites for the first monolayer, there is also no electrostatic problem. What we then have is that the first monolayer constitutes basically of a charge-neutral non-polar rocksalt FeO layer with 25% Fe vacancies. All Fe ions are occupying the B-site with 33% of them having the 2+ valence and 67% the 3+ state and the vacancies are not ordered since we did not observe any superstructure. We have also carried out polarization dependent XAS measurements, and we are able to indeed verify in detail that also the dichroic spectrum is consistent with the 33% $Fe^{2+}$ and 67% $Fe^{3+}$ B-site occupation. See Supplemental

Materials [31, Figs. S5]. Please note that these XAS spectra and the dichroism therein are very different from those of Fe atoms on MgO [34]. Moreover, capping this monolayer with a thick layer of MgO induces spectral weight changes that are fully consistent with the presence of 25% Fe vacancies. See Supplemental Materials [31, Fig. S6, 33].

For the second monolayer, we observe in the experiment the appearance of some amount of A-sites, about 16.7%, see top panel of Fig. 3. We now can arrive at the following model, see Fig. 4 where the left panel shows the growth process and the right panels the corresponding net charges, electric field, and electric potential of each plane. Since in bulk magnetite, a monolayer per unit cell includes 2 A-site $Fe^{3+}$, 2 B-site $Fe^{2+}$, 2 B-site $Fe^{3+}$, and 8 oxygen ions, we will use the formula notation $Fe_6O_8$ instead of $Fe_{3/4}O$ to describe each monolayer. When deposited, the second monolayer will first form a non-polar monolayer, like the first monolayer. Then, both the first and the second layers give away one $Fe^{3+}$ ion as shown in Fig. 4 (b) to the space in between them to form a layer with two A-site $Fe^{3+}$ ions. The entire film remains then also non-polar: a 6+ charge-layer is sandwiched by two 3- charge-layers. The electric field oscillates symmetrically around zero, and the electric potential remains nullified. In this model, there are two A-sites per 2 ML formula unit, i.e. the A-site concentration is $2/(2\times6) = 16\%$, which is very close to the experiment. The remaining outer $Fe_5O_8$ (defective rocksalt) layers contain each three $Fe^{3+}$ and two $Fe^{2+}$ B-site ions. The concentration of the $Fe^{2+}$ B-site ions is thus $(2\times2)/(2\times6) = 33\%$, i.e. the same as for the first monolayer and consistent with the experiment, see top panel of Fig. 3.

For a 3-ML film, the layer added will again form a non-polar monolayer first. This monolayer and the subsurface monolayer then carry out the same process in which both give away one $Fe^{3+}$ ion to the space in between. See Fig. 4 (c). Again, the potential divergence remains nullified after this process as shown on the most right panel of Fig. 4 (c). One complete bulk $Fe_6O_8$ layer now is formed. This growth process is repeated for the subsequent layers, and the model predicts that the concentration of the A-site ions will increase following the geometrical series $2(n-1)/6n$, while the concentration of the $Fe^{2+}$ B-site ions will remain constant at $2n/6n = 33\%$ and that of the $Fe^{3+}$ B-site ions will decrease following $2(n+1)/6n$, where n denotes the number of monolayers. These predictions of the model are also presented in Fig. 3. We can see that the essential behavior observed in the experiment is well reproduced with convergence to the bulk values for thicker films. We also would like to note that an ordering in the outer $Fe_5O_8$ layer can be made consistent with the often observed $(\sqrt{2}\times\sqrt{2})R45°$ surface reconstruction in thicker (001) $Fe_3O_4$ films, please see the Supplementary Materials for details [31].

We thus have found that the A-sites are absent in the first monolayer or interface, and that Fe ions are on the move while the films is growing to accommodate for the presence of A-sites inside the film having the proper crystal structure and stoichiometry. We clearly have a 'dynamic atomic reconstruction' taking place here.

It is interesting to note that the $Fe_3O_4$ thin films are insulating and that the interface does not induce metallicity as shown by the resistivity measurements displayed in Fig. 1 (f). This is obviously in contrast with the resistivity measurements on $SrTiO_3$/$LaAlO_3$ [5,35–37] and $SrTiO_3$/$RETiO_3$ [13–15]. In principle, the $Fe_5O_8$ interface layer could have been conducting since this layer can be considered as a defective and doped rocksalt FeO layer. Yet, considering the fact that small polaron effects in bulk $Fe_3O_4$ are strong and hamper the conductivity [38–40], we may expect that this will also be the case for the interface layer. Its resistivity will then be dominated by strong scattering effects due to disorder.

Our findings have direct and important implications for the field of $Fe_3O_4$ spintronics. There are some reports concerning the possible existence of a magnetically dead layer at the interface [41–43], but others ascribe the decrease of the magnetization in the thin films to the presence of antiphase boundaries leading to superparamagnetic behavior of the domains [44–46]. Our findings may give credit to the proponents of the dead magnetic layer model. In view of the absence or low amount of A-sites in the interface region, some of the superexchange paths which determine the ferrimagnetism in $Fe_3O_4$ are certainly missing. This then would also explain why tunneling experiments have spin-polarizations different than expected from the properties of bulk $Fe_3O_4$ [47]. We now can propose that the insertion of a monolayer of magnetic metals like Fe, Co, Ni, or even noble metals like Cu, Ag, Au or Pt between the $Fe_3O_4$ and the insulating oxide substrate will drastically change the situation: the metal layer inserted will act as a charge reservoir that can accommodate the flow of planar charges required to stabilize a $Fe_3O_4$ interface layer which has A-sites like in the bulk. The occurrence of a magnetically dead layer can then be prevented and also the spin polarization at the interface may be increased. A hint that the latter is not unrealistic can be found in an early work by Dedkov *et al.* [48] on oxidized Fe films deposited on metal substrates.

To summarize, using soft x-ray absorption spectroscopy we find that nature provides us with an unexpected but elegant solution for the polar catastrophe problem at the $Fe_3O_4$/MgO (001) interface: the A-site $Fe^{3+}$ ions are missing in the first $Fe_3O_4$ layer and the growth process involves movements of not only the surface but also the subsurface Fe ions, securing epitaxy and stoichiometry at the same time. Having identified this 'dynamic atomic reconstruction' growth principle, we conclude that we really have to think differently and openly about how polar interfaces can grow. Apparently, 'nature' offers us a much wider range of opportunities to prepare unstable polar interfaces. It would be interesting to put effort to grow a monolayer or a few monolayer of $Fe_3O_4$ film where the defects are ordered, so that diffraction techniques can confirm the growth model.

**Acknowledgments:**

The research in Köln was supported by the Deutsche Forschungsgemeinschaft through SFB 608. The research of X. H. L. was supported by the Max Planck-POSTECH Center for Complex Phase Materials.

$(\sqrt{2} \times \sqrt{2})$R45° surface reconstruction which is consistent with the $Fe_5O_8$ surface as proposed from our model.

**Figures:**

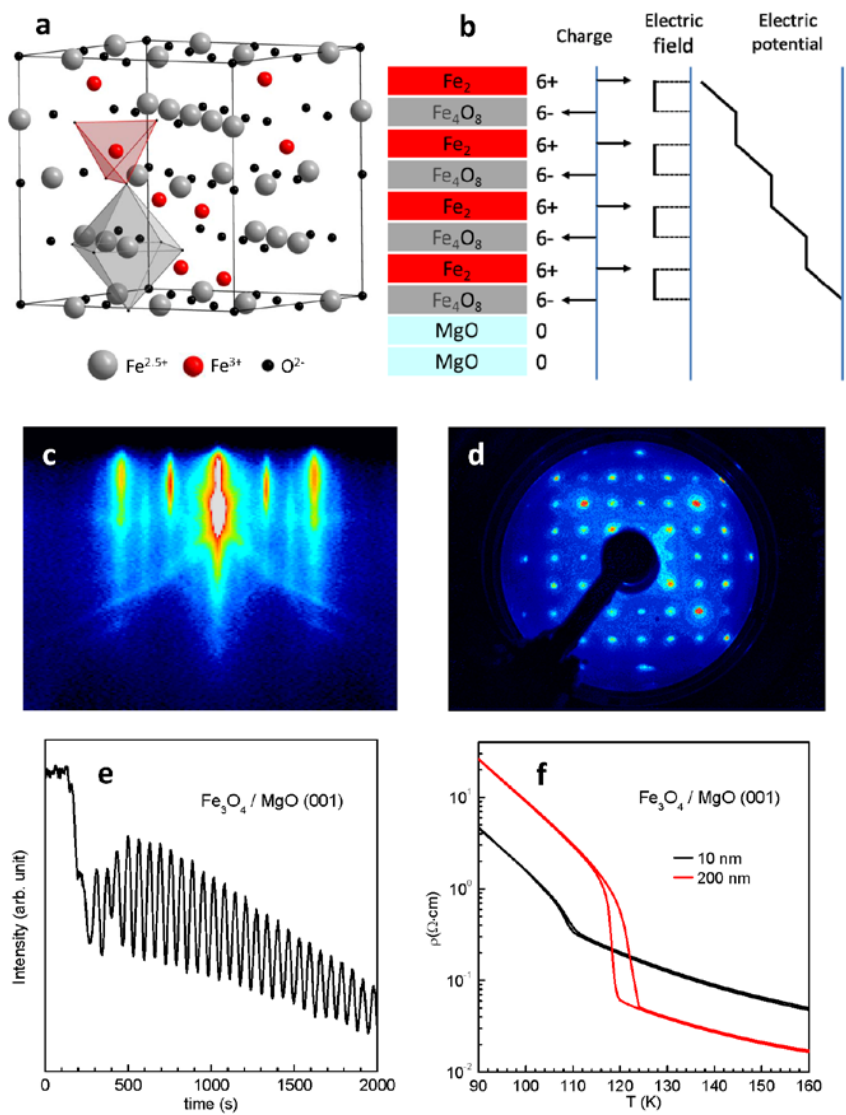

**Fig. 1. Fe₃O₄ on MgO (001).** (a) structure of Fe$_3$O$_4$. (b) build-up of the polar catastrophe of Fe$_3$O$_4$ on MgO (001): charged planes, and corresponding charge, electric field, and electric potential. (c) and (d) RHEED and LEED patterns, respectively, of a 200 nm epitaxial Fe$_3$O$_4$ (001) film showing the characteristic $(\sqrt{2} \times \sqrt{2})$R45° superstructure. (e) RHEED intensity oscillations of the specularly reflected beam. The electron beam was incident along the [100] direction, with a primary energy of 20 kV. (f) resistivity as a function of temperature of a 10 nm and a 200 nm thin film, showing the presence of the Verwey transition.

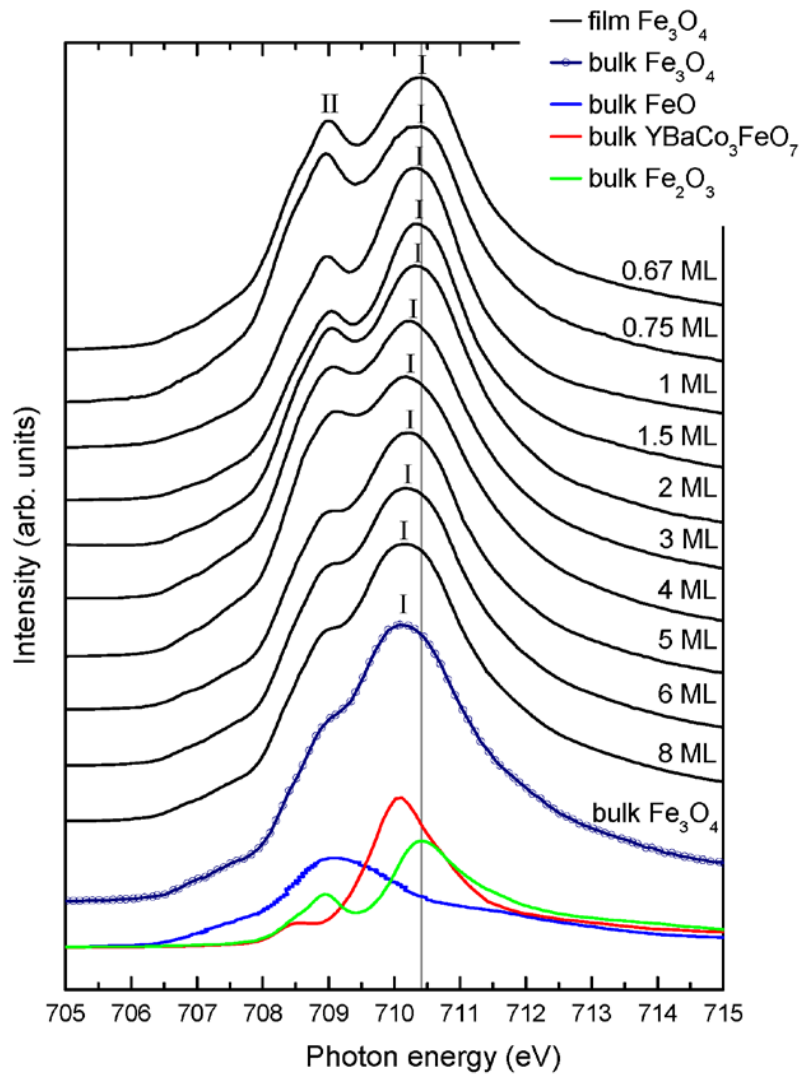

**Fig. 2. Fe $L_3$ XAS spectra of Fe$_3$O$_4$ films.** The film thickness varies from 0.67 to 8 ML. The reference spectra of bulk Fe$_3$O$_4$, bulk YBaCo$_3$FeO$_7$ (Fe$^{3+}$ in tetrahedral coordination) [28], bulk FeO (Fe$^{2+}$ in octahedral coordination) [32] and bulk Fe$_2$O$_3$ (Fe$^{3+}$ in octahedral coordination) are also included. All spectra were measured at 300 K. The gray line indicates the energy position of the main peak in the spectrum of bulk Fe$_2$O$_3$. The full Fe $L_{2,3}$ spectral range is presented in the Supplementary Materials.

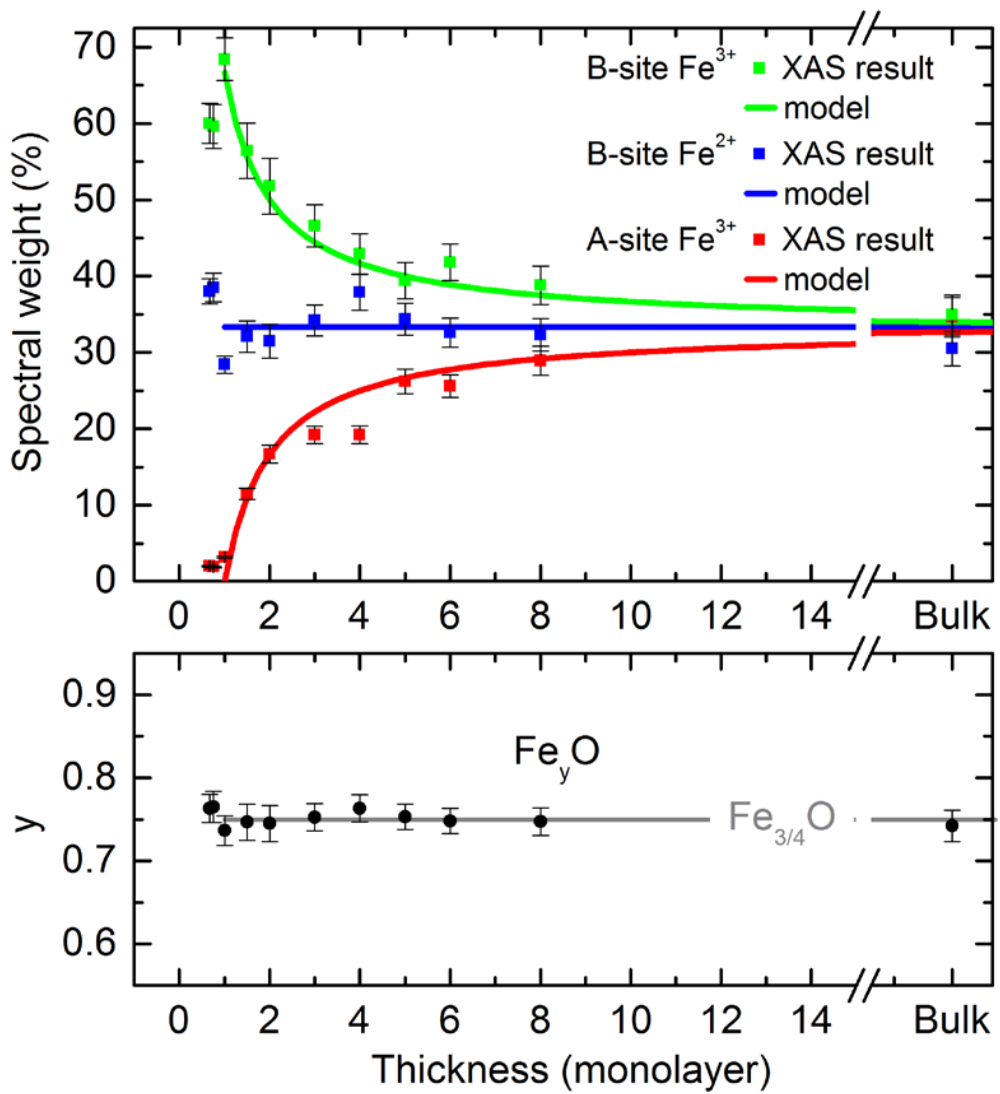

**Fig. 3. The extracted spectral weight of each Fe site versus the film thickness.** The spectral weight of the B-site $Fe^{3+}$, the B-site $Fe^{2+}$, and the A-site $Fe^{3+}$ are given by the green, blue, and red closed squares, respectively. Green, blue, and red curves depict the concentrations of the B-site $Fe^{3+}$, the B-site $Fe^{2+}$, and the A-site $Fe^{3+}$, respectively, in our model, see the text. The Fe content (y) of the $Fe_yO$ films derived from the average valence are shown in the bottom panel. The error bars reflect the deviations of the fits to the experimental data.

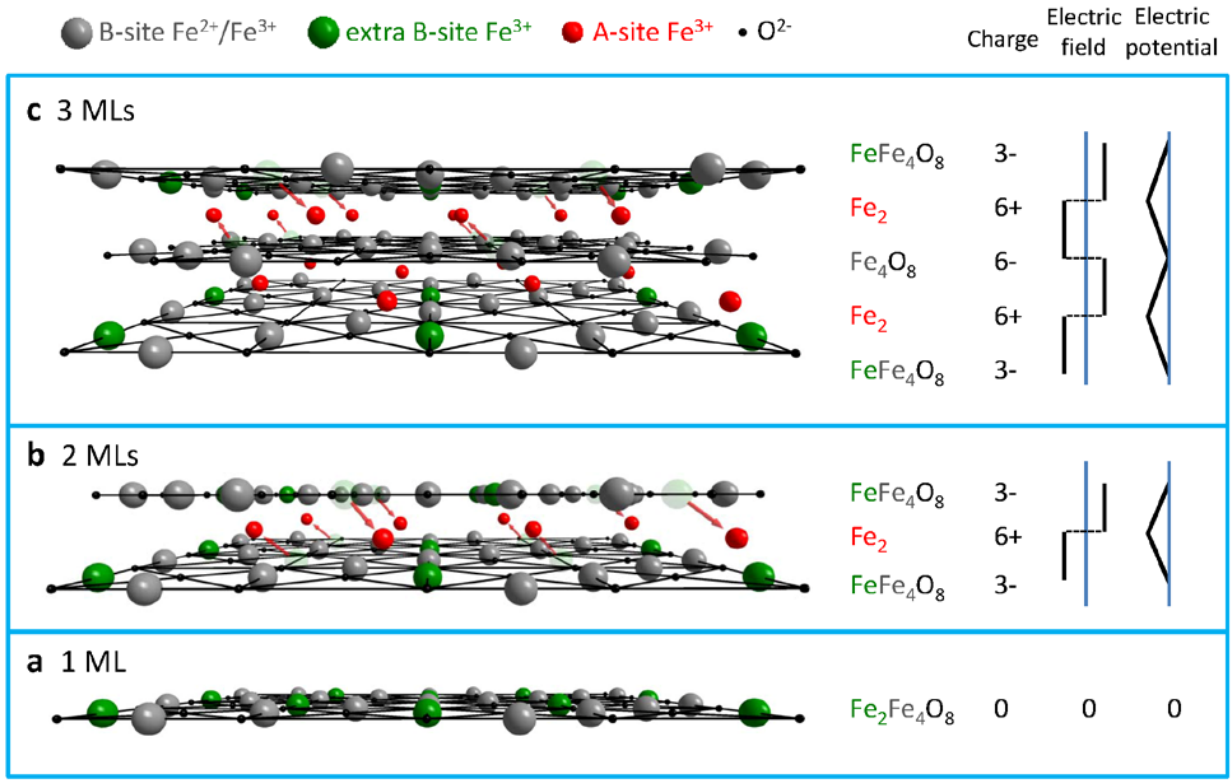

**Fig. 4. Model for the growth process of polar Fe₃O₄ (001) thin films.** (a) 1 monolayer, (b) 2 monolayers, and (c) 3 monolayers.

**Supplementary materials:**

# Dynamic atomic reconstruction:
# how $Fe_3O_4$ thin films evade polar catastrophe for epitaxy

C. F. Chang,[1] Z. Hu,[1] S. Klein,[2,†] X. H. Liu,[1] R. Sutarto,[2,†] A. Tanaka,[3] J. C. Cezar,[4,†]
N. B. Brookes,[4] H.-J. Lin,[5] H. H. Hsieh,[6] C. T. Chen,[5] A. D. Rata,[1,†] and L. H. Tjeng[1]

**Thin film growth and XAS measurement.** $Fe_3O_4$ thin films were epitaxially grown by molecular beam epitaxy (MBE). The base pressure of the MBE system was in the low $10^{-10}$ mbar range. High purity Fe metal was evaporated from an alumina crucible in a pure oxygen atmosphere of $3\times10^{-7}$ mbar onto clean and annealed MgO (001) substrates. The substrate temperature was kept at 250 °C during growth. *In-situ* and *real-time* monitoring of the epitaxial growth was performed by reflection high energy electron diffraction (RHEED). Oscillations in the RHEED specular beam intensity, where each oscillation corresponds to the formation of one new atomic monolayer (ML), allows for precise control of the film thickness. The RHEED oscillation measurements were done using cleaved MgO substrates, all other thin film results in this paper were obtained from epi-polished MgO substrates. The XAS measurements were performed at the 11 A Dragon beamline of the National Synchrotron Radiation Research Center (NSRRC) in Taiwan using *in-situ* MBE grown samples. The spectra were recorded at 300 K using the total electron yield method (TEY) in a chamber with a base pressure of $2\times10^{-10}$ mbar. The photon energy resolution at the Fe $L_{2,3}$ edges ($h\nu \sim$ 700–740 eV) was set at 0.3 eV and the degree of linear polarization was 99%. The samples were mounted on a holder which was tilted with respect to the incoming beam, such that the Poynting vector of the light makes an angle of 70° with respect to the [001] surface normal. The angle ($\theta$) between the electric field vector $E$ and the [001] surface normal can be varied between 20° and 90°. $I_{ip}$ and $I_\perp$ are the spectra measured at $\theta$ = 20° and 90°, respectively. The isotropic XAS spectra can be obtained via the formula of $I = I_\parallel + 2I_\perp$, where $I_\parallel$ is the spectrum with the $E$ parallel to the [001] surface normal extracted from $I_\parallel = [I_{ip} - I_\perp \cos^2(70°)]/\sin^2(70°)$. Magnetic circular dichroism (MCD) spectra of bulk magnetite crystals were measured at the ID8 beamline of the European Synchrotron Radiation Facility (ESRF) in Grenoble. The magnetic field was set at 5 Tesla. Fig. S1 shows the full Fe $L_{2,3}$ XAS spectra of the $Fe_3O_4$ films together with the spectra of bulk $Fe_3O_4$, bulk $YBaCo_3FeO_7$ ($Fe^{3+}$ in tetrahedral coordination) [1], bulk FeO ($Fe^{2+}$ in octahedral coordination) [2] and bulk $Fe_2O_3$ ($Fe^{3+}$ in octahedral coordination): the spectra are identical to those in Figure 2 in the main text, but with a wider photon energy window covering both Fe $L_3$ and $L_2$ edges.

**Configuration interaction cluster calculation.** To interpret and better understand the x-ray absorption (XAS) spectra and their thickness dependence we have performed simulations using the well established configuration interaction cluster model that includes the full atomic multiplet theory and the local effects of the solid [3-5]. It accounts for the intra-atomic Fe $3d$–$3d$ and $2p$–$3d$ Coulomb and exchange interactions, the atomic $2p$ and $3d$ spin-orbit couplings, the O $2p$–Fe $3d$ hybridization and the local ionic crystal field. The calculations were done using the program XTLS 8.3 [5]. The XAS spectra of $Fe_3O_4$ can be decomposed into the three sub-spectra of the three Fe sites, *i.e.* A-site $Fe^{3+}$, B-site $Fe^{2+}$, and B-site $Fe^{3+}$. We have considered an $FeO_4$ and an $FeO_6$ cluster for each Fe A-site and B-site, respectively. Parameters for the multipole part of the Coulomb interactions were given by 75% and 80% of the Hartree-Fock values for the $d-d$ and $p-d$ Slater integrals, respectively, while the monopole parts ($U_{dd}$, $U_{cd}$) as well as the O $2p$–Fe $3d$ charge transfer energy ($\Delta$) were adopted from typical values for $Fe^{2+}$ and $Fe^{3+}$ ions [6,7]. The hopping integrals between the Fe $3d$ and O $2p$ were calculated for the various Fe–O bond lengths according to Harrison's description [8]. The Fe–O bond lengths were taken from x-ray single-crystal structure diffraction data [9]. The crystal field parameter $10Dq$ was tuned to fit the experimental spectra. All parameters are listed in Ref. 10.

Figure S2 shows the experimental Fe $L_{2,3}$ XAS (open circle) and magnetic circular dichroism (MCD, open diamond) spectrum of bulk $Fe_3O_4$ together with the simulation results (magenta and purple lines, respectively). The three components, the sum of circular polarization spectra of each site, *i.e.* B-site $Fe^{2+}$ (blue line), A-site $Fe^{3+}$ (red line), and B-site $Fe^{3+}$ (green line) are also depicted. The relative energy positions for the three sub-spectra were determined in such a way that the simulated total MCD spectrum fits by the experimental MCD spectrum best, see Refs. 6, 11, and 12. The fits were done using the "NMimimize" function of the Mathematica software [13].

By making weighted sums with the three isotropic sub-spectra using the "NMimimize" function of the Mathematica software [13] to obtain the best fit to the experimental spectrum of each Fe film with the different thicknesses, we extract the relative amount of B-site $Fe^{2+}$, A-site $Fe^{3+}$, and B-site $Fe^{3+}$ ions as a function of film thickness. The XAS simulation results of the $Fe_3O_4$ thin films of 0.67, 0.75, 1, 1.5, 2, 3, 4, 5, 6, and 8 MLs are shown in Fig. S3.

To double check the validity of the three isotropic sub-spectra, we compare them in Figure S4 with the experimental XAS spectra of the standard references for each Fe site, i.e., bulk $YBaCo_3FeO_7$ [1] for the A-site $Fe^{3+}$, bulk FeO [2] for the B-site $Fe^{2+}$, and bulk $Fe_2O_3$ for the B-site $Fe^{3+}$ (same as those shown in Figure 2 in the main text, and in Figure S1 in the supplementary materials). Each of them reproduces the experimental spectrum of its corresponding reference very well. We include also in Figure S4 the XAS spectrum from $Fe_{0.04}Mg_{0.96}O$ [14], a system of Fe impurities embedded in MgO. The identical spectral features of the bulk FeO and of the Fe impurity system clearly demonstrate that XAS is most sensitive to the presence of the nearest neighbor ligands only. We have also done calculations for an $Fe^{2+}$ in $FeO_5$ and an $Fe^{3+}$ in $FeO_5$ by simply removing the apical oxygen of the $FeO_6$, but otherwise using the same parameters as those for $FeO_6$, as shown in Figure S4. Only minor differences can be observed between the isotropic XAS spectra of an $Fe^{2+}$ in $FeO_6$ and in $FeO_5$, and of an $Fe^{3+}$ in $FeO_6$ and $FeO_5$. The large difference between the isotropic XAS spectra of octahedral $Fe^{3+}$ and tetrahedral $Fe^{3+}$ originates from the fact that the effective 10Dq ligand/crystal field value is positive for the octahedral coordination while it is negative for the tetrahedral coordination.

**1 ML $Fe_3O_4$ thin film: polarization dependence.** Figure S5 shows the experimental linear polarization-dependent Fe $L_{2,3}$ XAS spectra of the 1 ML $Fe_3O_4$. In the bottom panel, the experimental linear dichroic (LD) signal, defined as the difference between two polarizations (E ∥ C – E ⊥ C) is displayed, together with the calculated LD spectrum for the scenario of 33 % B-site $Fe^{2+}$ and 67 % B-site $Fe^{3+}$. The LD signal can be well reproduced without including any contribution from the A-site $Fe^{3+}$ ion. All this can be very well understood: the isotropic spectrum is determined mostly by the octahedral part of the ligand/crystal field, while the dichroism is due to the small tetragonal part of the crystal field in the monolayer. This tetragonal part of the crystal field makes the orbital occupation of the high-spin $d^6$ ion to become anisotropic, resulting in the polarization dependence of the intensity of the $Fe^{2+}$ signal. The tetragonal part of the crystal field does not affect the orbital occupation of the spherical high-spin $d^5$ ion but sets up the energy splitting in the XAS final states, resulting in the polarization dependence of the $Fe^{3+}$ peak position. Please note that these XAS spectra and the dichroism therein are very different from those of Fe atoms on MgO [15], confirming the notion that $L_{2,3}$-XAS is indeed an extremely powerful method to determine the local electronic structure of transition metal systems.

**1 ML $Fe_3O_4$ thin film capped with 10 ML MgO.** Fig. S6 shows the Fe $L_{2,3}$-XAS spectra of the 1 ML $Fe_3O_4$ film, the 1 ML $Fe_3O_4$ film capped with 10 ML MgO, and the $Fe_{0.04}Mg_{0.96}O$ system [14]. One can clearly observe that the spectrum of the 1 ML $Fe_3O_4$ changes drastically upon capping with MgO and that the spectrum becomes identical to that of octahedral $Fe^{2+}$ in $Fe_{0.04}Mg_{0.96}O$. This means that the presence of Mg converts all available $Fe^{3+}$ into $Fe^{2+}$, indicating that the 1 ML $Fe^{2+}Fe^{3+}Fe^{3+}O_4$ film is reacted to 1 ML $Mg^{2+}Fe^{2+}Fe^{2+}Fe^{2+}O_4$. This in turn establishes that the 1 ML $Fe_3O_4$ film is indeed an $Fe_{0.75}O$ monolayer, i.e. an FeO monolayer with 25% Fe vacancies

**($\sqrt{2} \times \sqrt{2}$)R45° surface reconstruction.** The $Fe_5O_8$ surface as proposed from our model is consistent with the $(\sqrt{2} \times \sqrt{2})$R45° superstructure of the surface as shown in Fig. S7. Starting with the nonpolar $Fe_6O_8$ monolayer, all Fe ions are in an octahedral coordination. It consists of a bulk $Fe_4O_8$ layer and two extra octahedral Fe cations. Every other extra octahedral Fe cation is then removed to help to form the A-site layer below. The remaining extra octahedral Fe ions in the $Fe_5O_8$ surface layer can then be arranged as to give the $(\sqrt{2} \times \sqrt{2})$R45° superstructure. In Fig. S7 we also have included the directions (arrows) of the atomic relaxations forming a wavelike surface patterns along the [110].

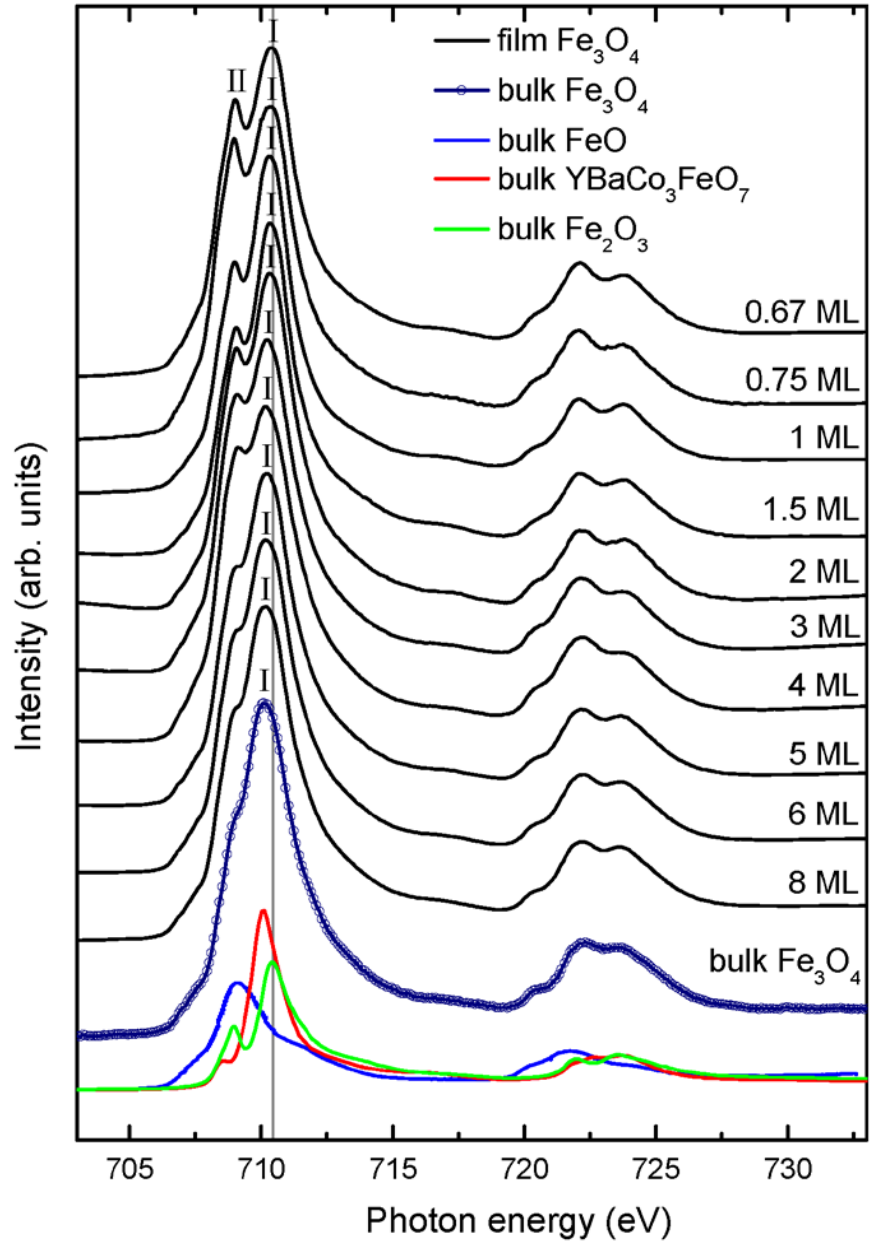

**Figure S1.** The full range of the Fe $L_{2,3}$ XAS spectra depicted in Fig. 2.

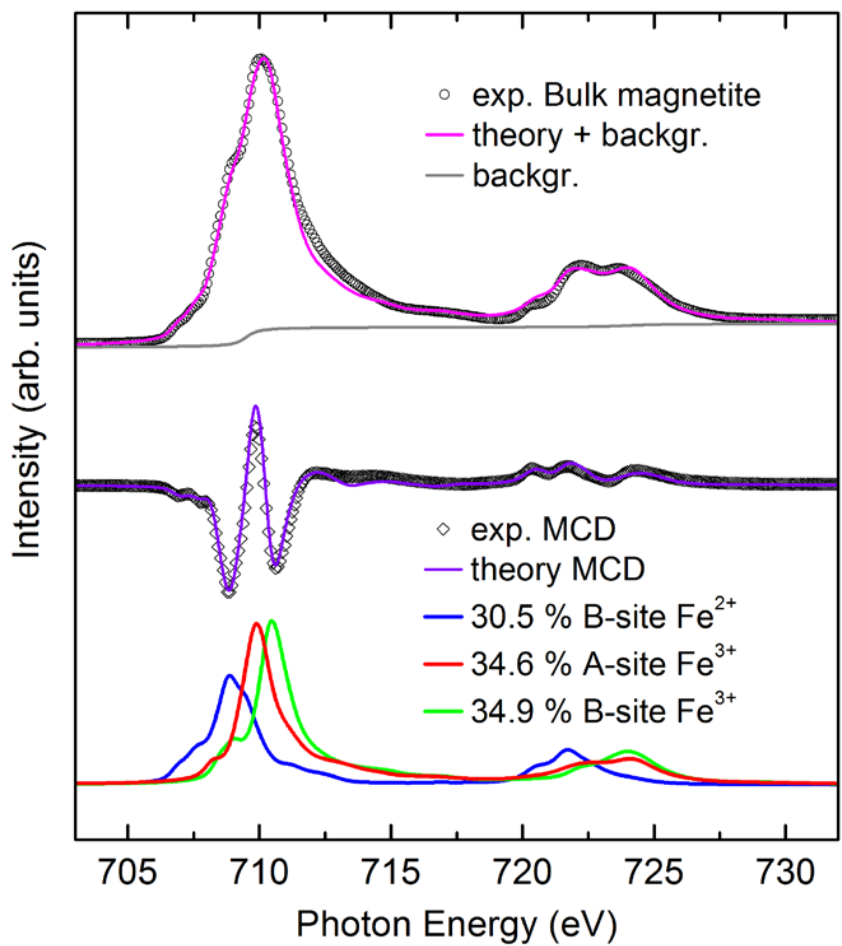

**Figure S2.** Fe $L_{2,3}$ XAS and MCD spectra of bulk $Fe_3O_4$. The XAS (open circle) and MCD spectra (open diamond) were taken at 300 K. Magenta and purple curves are the simulated XAS and MCD spectra, respectively. The gray line is an arctangent like XAS background. Simulations for the three components are displayed in the bottom panel: B-site $Fe^{2+}$ (blue line), A-site $Fe^{3+}$ (red line), and B-site $Fe^{3+}$ (green line).

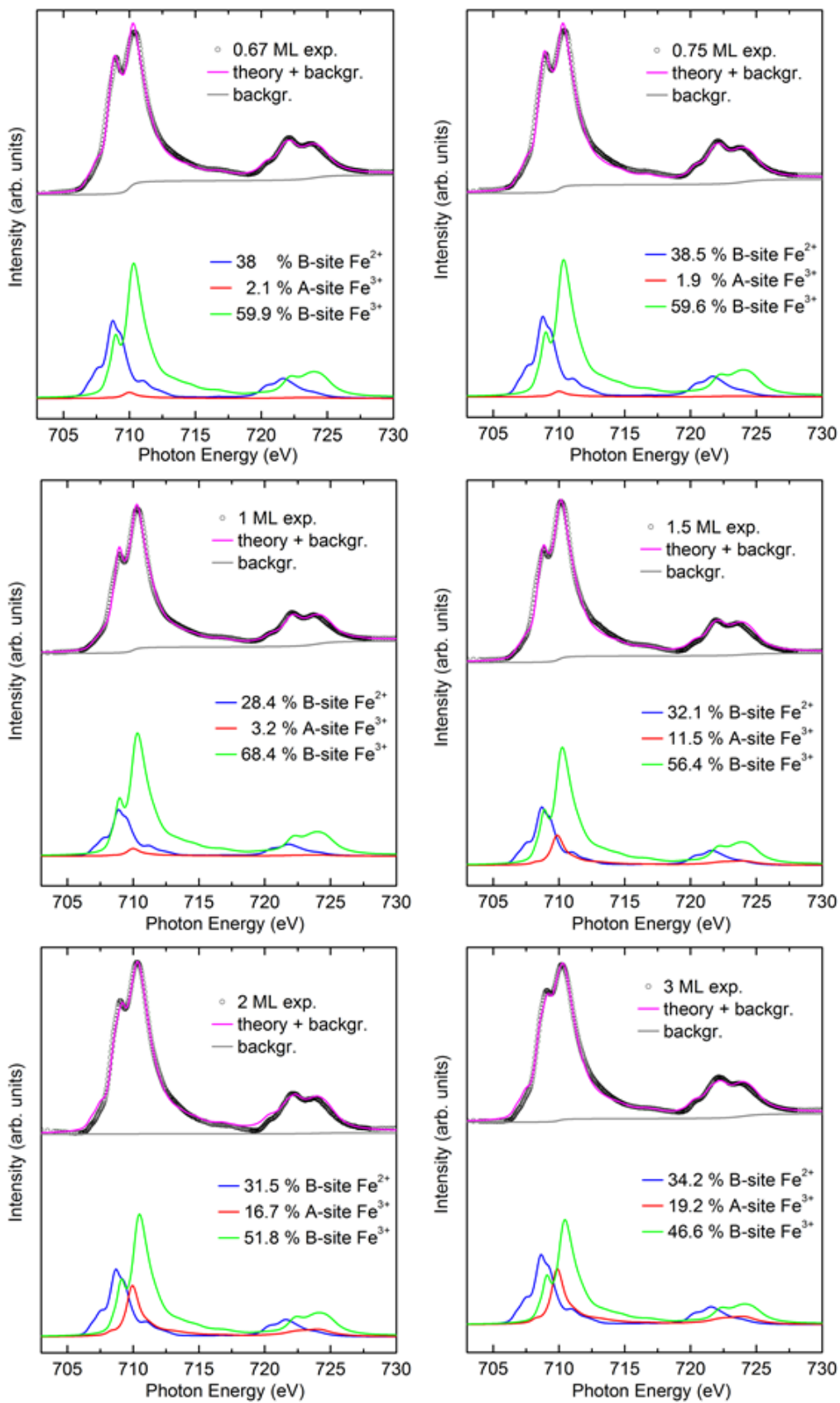

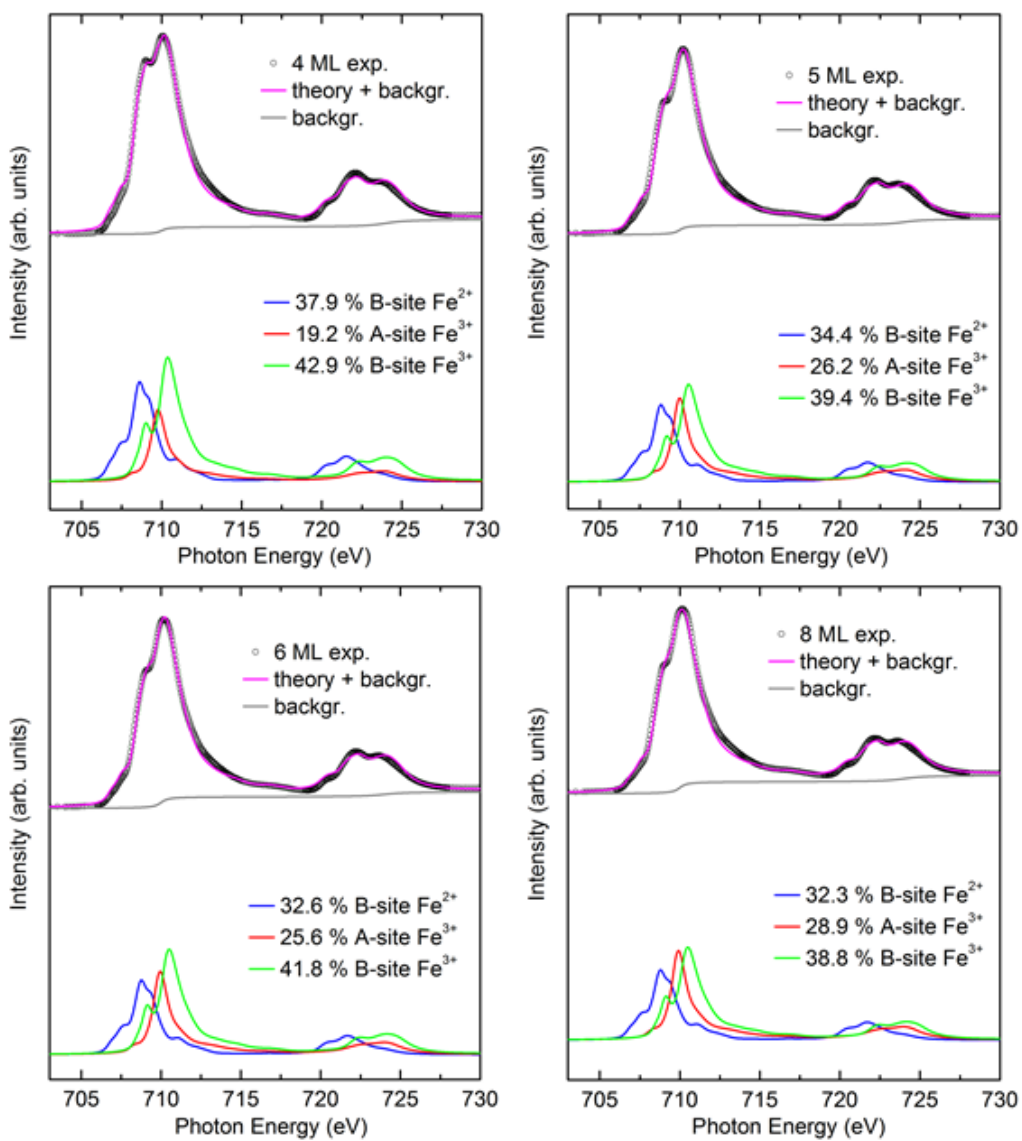

**Figure S3.** The XAS simulation results of the $Fe_3O_4$ thin films of 0.67, 0.75, 1, 1.5, 2, 3, 4, 5, 6, and 8 MLs. The experimental XAS (open circle) spectra are those in Fig. S1. The magenta curve is the simulated XAS spectrum. The gray line is an arctangent like XAS background. Simulations for the three components are displayed in the bottom panel: B-site $Fe^{2+}$ (blue line), A-site $Fe^{3+}$ (red line), and B-site $Fe^{3+}$ (green line).

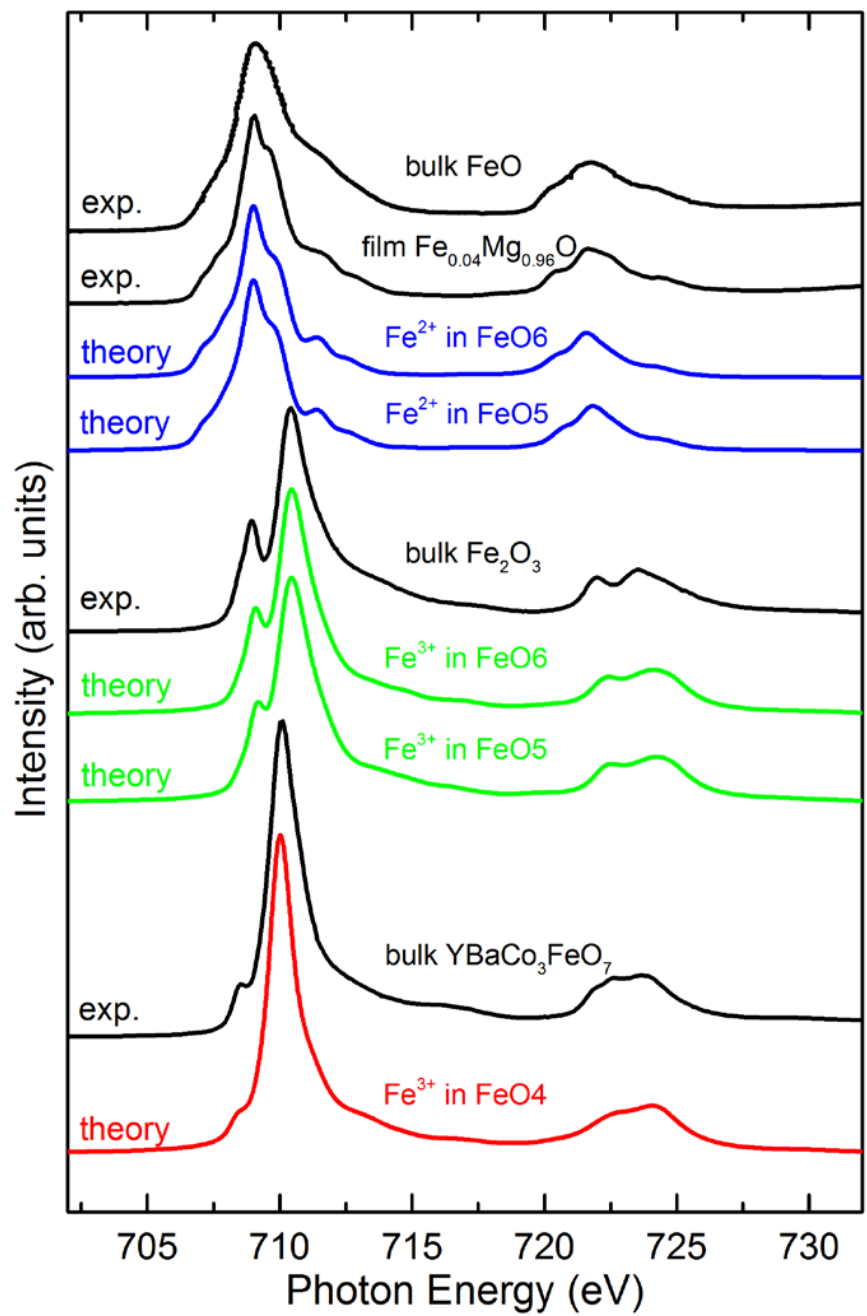

**Figure S4.** Experimental Fe $L_{2,3}$ XAS spectra of bulk FeO [2] ($Fe^{2+}$ in octahedral coordination), thin film $Fe_{0.04}Mg_{0.96}O$ [13] ($Fe^{2+}$ in octahedral FeO6), bulk $Fe_2O_3$ ($Fe^{3+}$ in octahedral FeO6), and bulk $YBaCo_3FeO_7$ [1] ($Fe^{3+}$ in tetrahedral FeO4), together with the simulated spectra of an $Fe^{2+}$ in FeO6 and in FeO5, an $Fe^{3+}$ in FeO6 and FeO5, and an $Fe^{3+}$ in FeO4.

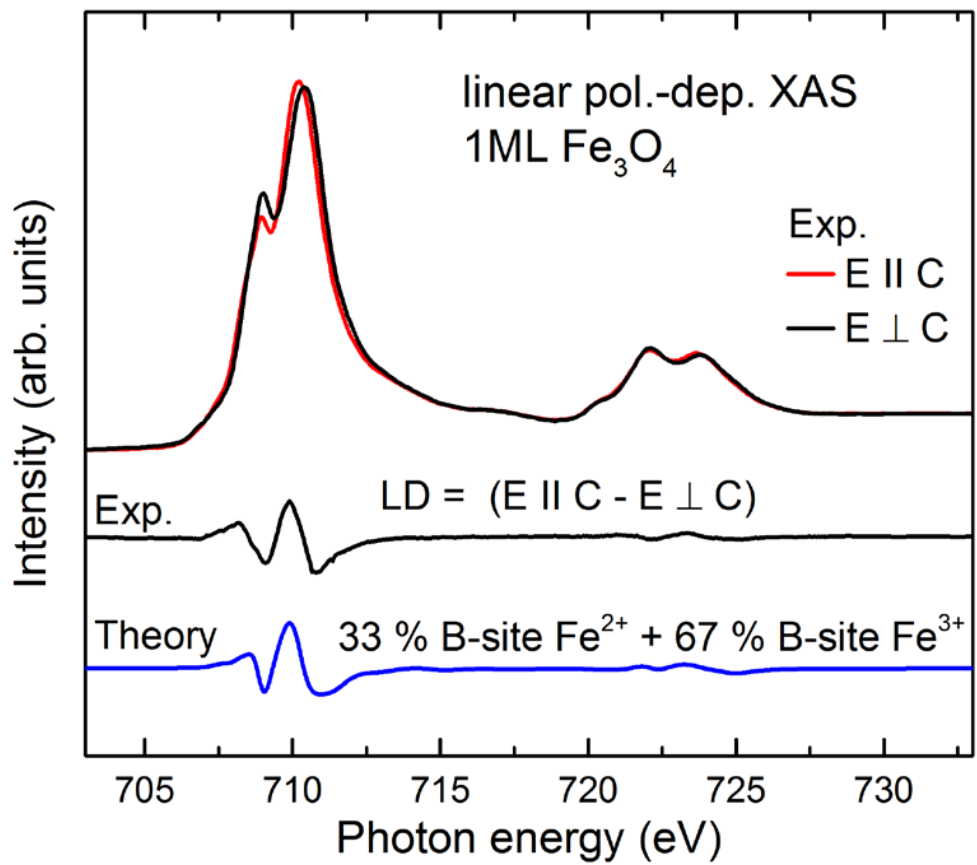

**Figure S5.** Experimental linear polarization-dependent Fe $L_{2,3}$ XAS spectra of 1 ML $Fe_3O_4$. The experimental linear dichroic (LD) signal defined as the difference between two polarizations (E ∥ C − E ⊥ C) is shown in the middle, together with the calculated LD spectrum for the scenario of 33 % B-site $Fe^{2+}$ and 67 % B-site $Fe^{3+}$ (no A-site $Fe^{3+}$).

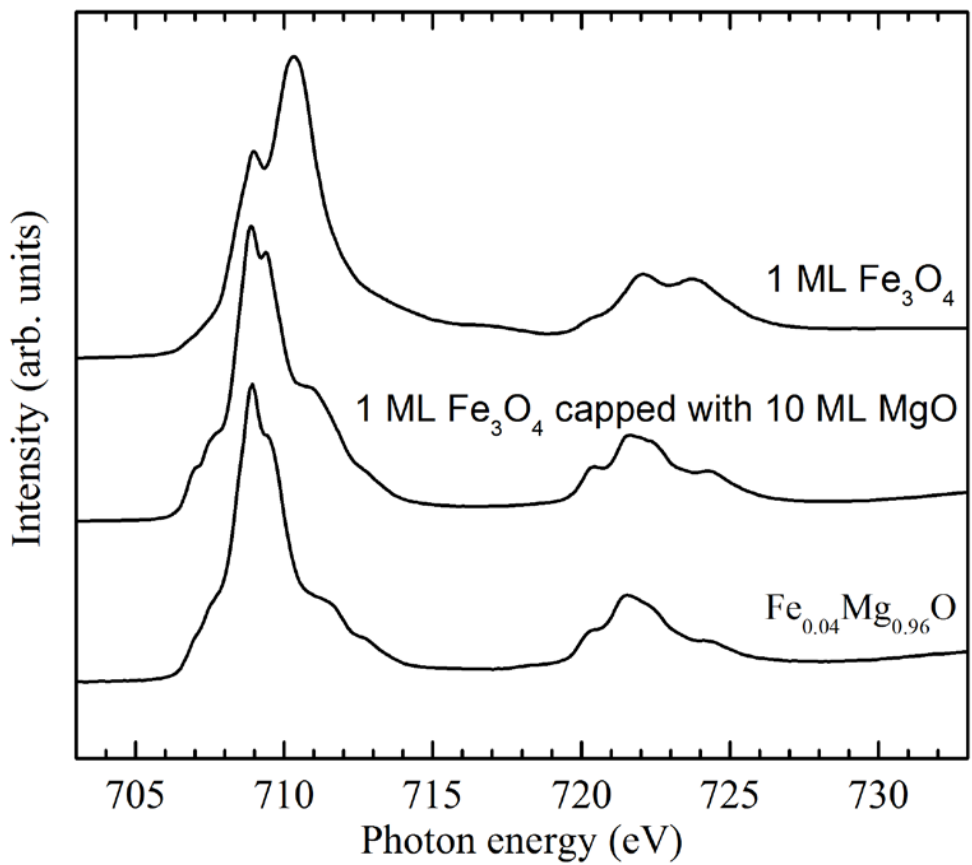

**Figure S6.** Fe $L_{2,3}$ XAS spectra of the 1 ML Fe$_3$O$_4$, the 1 ML Fe$_3$O$_4$ capped with 10 ML MgO, and Fe$_{0.04}$Mg$_{0.96}$O thin film.

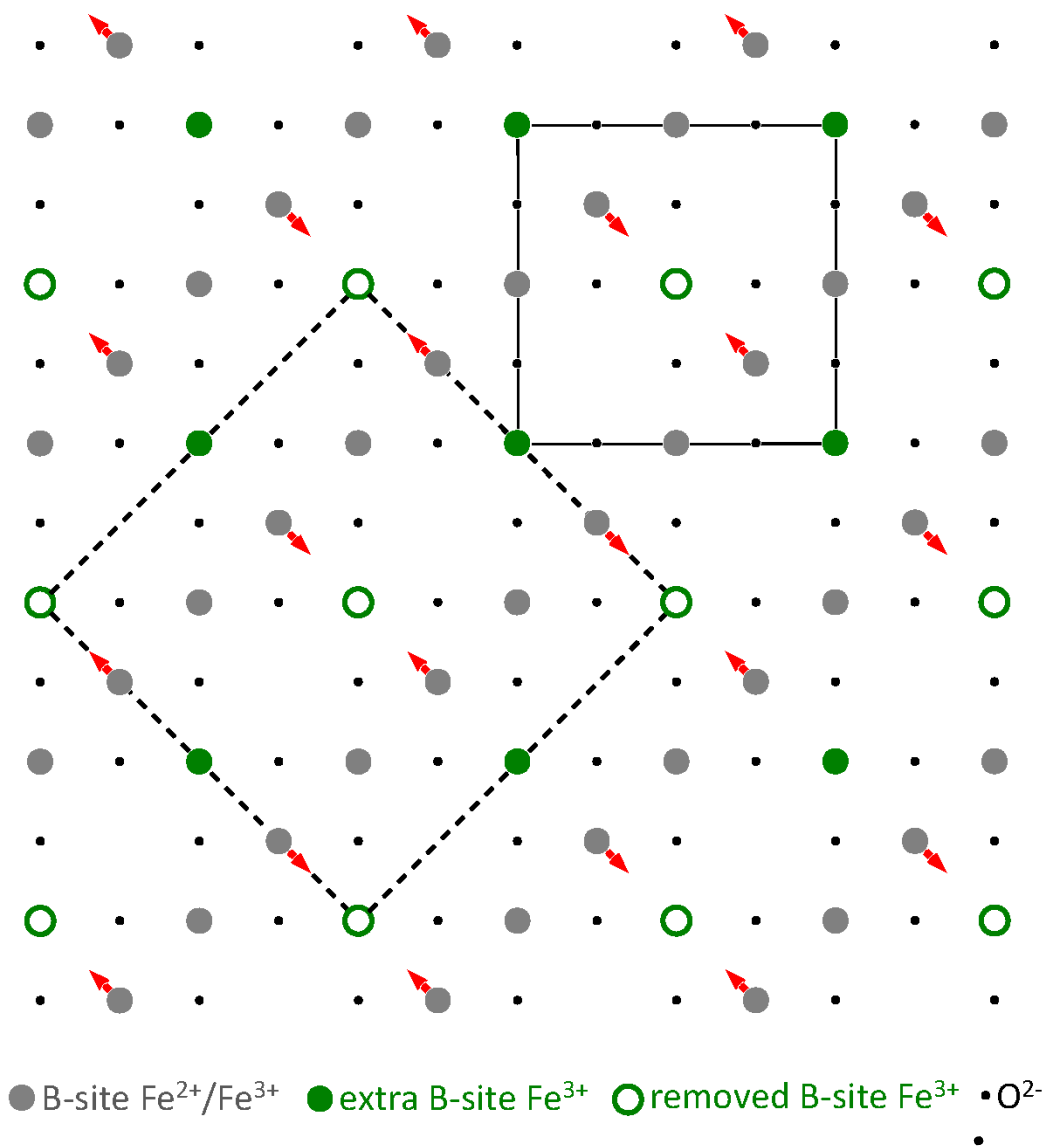

**Figure S7.** Proposed $(\sqrt{2} \times \sqrt{2})$R45° surface reconstruction of $Fe_3O_4$ (001). The solid-line square is the conventional cubic unit cell. The dashed-line square is the superstructure unit cell. Also included are the directions (arrows) of the atomic relaxations forming a wavelike surface patterns along the [110].

# References of the Supplementary materials